\documentclass[12pt]{aastex}

\newcommand{\be}{\begin{equation}}
\newcommand{\ee}{\end{equation}}
 
\newcommand{\mtot}{m_{\rm tot}}

\begin{document}
\title{Extrasolar Trojans: The Viability and Detectability of Planets in the 1:1 Resonance} 

\bigskip
\author{Gregory Laughlin$^1$, John E. Chambers$^2$} 

\bigskip
\affil{$^1$ UCO/Lick Observatory, UCSC, Santa Cruz, CA 95064
laugh@ucolick.org}
\affil{$^2$ NASA Ames Research Center, Moffett Field, CA 94035
john@mycenae.arc.nasa.gov}

\begin{abstract} 

We explore the possibility that extrasolar planets might be found in the
1:1 mean-motion resonance, in which a pair of planets share a time-averaged
orbital period. There are a variety of stable co-orbital configurations, and
we specifically examine three different versions of the 1:1 resonance.
In the first configuration, the two planets and the star participate
in tadpole-type librations about the vertices of an equilateral triangle.
The dynamics of this situation resemble the orbits of Jupiter's Trojan
asteroids. We show analytically that an equilateral configuration consisting of a
star and two equal-mass planets is linearly
stable for mass ratios $\mu=2m_{pl}/(2m_{pl}+M_{\star})<0.03812$. When the equilateral
configuration 
is subjected to larger perturbations, a related 1:1 resonance occurs.
In this second family of 
configurations, the planet pair executes horseshoe-type orbits
in which the librating
motion in the co-rotating frame is symmetric about a $180^{o}$
separation. The Saturnian satellites Janus and Epimetheus provide 
a solar system example of this phenomena. In the case of equal mass
planets, a numerical survey indicates that
horseshoe configurations are stable over long periods for mass ratios
$\mu<0.0004$, indicating that a pair of Saturn-mass planets can
exist in this resonance. The third configuration which
we examine is more exotic, and involves a pair of planets which exchange
angular momentum in a manner that allows them to indefinitely
avoid close encounters. An illustrative example of this resonance 
occurs when one planet has a highly eccentric orbit while the other
planet moves on a nearly circular orbit; the periapses are in alignment,
and conjunctions occur near periapse. All three of these resonant
configurations can be stable over timescales comparable to or longer than
stellar lifetimes. We show that pairs of planets in 1:1 resonance
yield characteristic radial velocity signatures
which are not prone to the $\sin (i)$ degeneracy. Indeed, Keplerian fits to the 
radial velocities cannot reveal the presence of two planets in the 1:1
resonance. We discuss a dynamical fitting method for such systems, and 
illustrate its use with a simulated data set. Finally, we argue that
hydrodynamical simulations
and torqued three-body simulations indicate that 1:1 resonant pairs might
readily form and migrate within protostellar disks.

\end{abstract}


\section{Introduction} 

In 1772, Lagrange proved the existence of five equilibrium points in the
equilateral three-body problem (Lagrange 1867-1892).
A century later, in 1875, Routh described the librating motion of a particle
in the vicinity of the triangular equilibrium points $L_{4}$, and $L_{5}$
(e.g. Routh 1898). Near the turn of the last century, in 1906,
Max Wolf of the Heidelberg observatory
discovered the first Trojan asteroid, 588 Achilles. This object is
participating in stable librations about Jupiter's $L_{4}$ point.
The extent of the Jovian Trojan population is now thought to rival the
main asteroid belt (Shoemaker, Shoemaker, \& Wolfe 1989).

Several thousand nearby stars are now being surveyed for periodic
radial velocity variations which indicate the presence of extrasolar planets
(Marcy, Cochran \& Mayor 2000). 
Systems with more than one planet are starting to emerge from the surveys, and
among these multiple-planet systems there are already interesting examples of
resonant relationships. For example, the planetary system orbiting
$\upsilon$ Andromedae displays a secular resonance in which the arguments
of periastron of the outer two planets librate about an aligned configuration
(Chiang, Trebachnik \& Tremaine 2001), and the pair of planets orbiting
GJ 876 are clearly participating in the 2:1 resonance (Marcy et al 2001,
Laughlin \& Chambers 2001, Lee \& Peale 2001).

Given that resonances are likely to be very common among extrasolar planets,
one can ask whether is is possible to have co-orbital planets, both with masses
of the order of Jupiter. In this paper, we argue that such ``trojan planets'' are indeed
a viable possibility. Configurations of this type are stable over a fairly wide
and varied
range of orbital parameters. Furthermore, a 1:1 resonance between a pair of extrasolar
planets induces a characteristic radial velocity signature in the parent star.
Such a signature may already exist unrecognized in the radial velocity data set,
as the identification of a 1:1 resonance requires a self-consistent dynamical
fitting procedure. 

This paper is organized as follows. In the next section (\S 2), we describe the
dynamics of three flavors of 1:1 resonance. In \S 3 we show how self-consistent
fits to the radial velocities induced by co-orbital planets are derived. In
\S 4 we conclude with a discussion of how co-orbital planets might form.

\section{The Dynamics of the 1:1 resonance} 

It is well known that an equilateral triangle configuration represents
an equilibrium point in the restricted 3-body problem in which one
of the bodies has a negligible mass. The configuration is stable to small
perturbations provided that the smaller of the two massive
bodies has a mass less than $0.03852\ldots$ in units of the total mass
({\it e.g.} Murray and Dermott 1999). Here we examine the stability of
the equilateral triangle configuration in the general 3-body problem.

Consider 3 spherical bodies with masses $m_0$, $m_1$ and
$m_2$, such that $m_1,m_2\le m_0$,
confined to move in a plane. In a coordinate system
centered on $m_0$, the acceleration on body $m_1$ is given by
\begin{equation}
\ddot{\bf r}_1=-\frac{G(m_0+m_1)}{r_1^3}{\bf r}_1
+\frac{Gm_2}{\Delta^3}{\bf \Delta}-\frac{Gm_2}{r_2^3}{\bf r}_2
\end{equation}
where ${\bf r}_1$ is the vector from $m_0$ to $m_1$, ${\bf r}_2$ is the
vector from $m_0$ to $m_2$, ${\bf \Delta}$ is the vector from $m_1$
to $m_2$, and $\dot{x}=dx/dt$ etc.

Resolving the acceleration on $m_1$ into components in the radial and
tangential directions, and noting that the radial and tangential
accelerations can be expressed as $\ddot{r}_1-r_1(\dot{\theta}_1)^2$
and $r_1\ddot{\theta}_1+2\dot{r}_1\dot{\theta}_1$ respectively, we get
\begin{eqnarray}
\ddot{r}_1-r_1(\dot{\theta}_1)^2&=&
-G\left[\frac{(m_0+m_1)}{r_1^3}
+\frac{m_2}{\Delta^3}\right]r_1
+Gm_2\left[\frac{1}{\Delta^3}
-\frac{1}{r_2^3}\right]r_2\cos(\theta_1-\theta_2) \nonumber \\
r_1\ddot{\theta}_1+2\dot{r}_1\dot{\theta}_1&=&
-Gm_2\left[\frac{1}{\Delta^3}
-\frac{1}{r_2^3}\right]r_2\sin(\theta_1-\theta_2)
\label{eq2}
\end{eqnarray}
where $(r_1,\theta_1)$ are the polar coordinates of body $m_1$
with respect to $m_0$. The acceleration components for body $m_2$
have a similar form to those for $m_1$.

We now consider a configuration in which the 3 bodies lie at the
corners of an equilateral triangle of side $a$, with all the masses
revolving on circular orbits about the centre of mass with angular
speed $n$.  In this case $r_i=\Delta=a$,
$\dot{r}_i=\ddot{r}_i=\ddot{\theta}_i=0$ and $\dot{\theta}_i=n$ for
$i=1,2$, and Eqns.~\ref{eq2} are satisfied provided that
$n^2a^3=G\mtot$, where $\mtot$ is the sum of all the masses.  Hence,
an equilateral triangle configuration is an equilibrium point in the
general 3-body problem as well as in the restricted problem.

To examine the stability of configurations close to the equilibrium point,
we expand Eqns.~\ref{eq2} in terms of $r_i$ and $\theta_i$, being careful
not to make any assumptions about the relative sizes of the masses
at this stage. In expanding the equations, we define 2 new variables
$\theta=\theta_1-\theta_2$ and $s=(r_1-r_2)/a$, and we
consider small deviations from the equilibrium configuration
corresponding to $\theta=\pi/3$ radians and $s=0$. The expanded equations
for $m_1$ are

\begin{eqnarray}
\ddot{r}_1-2an(\dot{\theta}_1-n)-n^2r_1&=&
-\frac{G\mtot}{a^2}\left(3-\frac{2r_1}{a}\right) \nonumber \\
&-&\frac{Gm_2}{a^2}\left[B(\theta)(1-\cos\theta)+
\frac{3s}{2}(1+\cos\theta)\right] \nonumber \\
a\ddot{\theta}_1+2n\dot{r}_1&=&
-\frac{Gm_2\sin\theta}{a^2}
\left[B(\theta)-\frac{3s}{2}\right]
\end{eqnarray}
where
\begin{equation}
B(\theta)=\left[\frac{1}{2\sin(\theta/2)}\right]^3-1
=-\frac{3\surd 3}{2}\left(\theta-\frac{\pi}{3}\right)+\cdots
\end{equation}
The equations for $m_2$ are similar except that $\sin\theta$ is
replaced with $-\sin\theta$ and $s$ is replaced by $-s$. In Eqn. 3, we have
neglected terms proportional to ${s}{B(\theta)}$. These terms
are negligible for orbits very close to the equilibrium point.

We can obtain a single pair of equations by taking the difference of
the radial equations for $m_1$ and $m_2$, and also the difference of
the tangential equations. Introducing a new variable
$\alpha=\theta-\pi/3$, and retaining only the linear term in the
expansion for $B(\alpha)$, we get
\begin{eqnarray}
\ddot{s}-2n\dot{\alpha}+\Gamma n^2\alpha-(3-K)n^2s=0
\nonumber \\
\ddot{\alpha}+2n\dot{s}-n^2K\alpha+\Gamma n^2s=0
\label{eq5}
\end{eqnarray}
where $K$ and $\Gamma$ are constants that depend only on the masses:
\begin{eqnarray}
K&=&\frac{9(m_1+m_2)}{4\mtot} \nonumber \\
\Gamma&=&\frac{3\surd 3(m_1-m_2)}{4\mtot}
\end{eqnarray}

Following Murray and Dermott (1999), the general solution to
Eqn.~\ref{eq5} is $\alpha=\sum_{j=1}^4\alpha_j\exp(\lambda t)$ and
$s=\sum_{j=1}^4s_j\exp(\lambda t)$, where the $\alpha_j$ and $s_j$ are
constants. The frequencies $\lambda$ are found by solving the determinant
of the 4 first-order equations equivalent to the two second-order
equations in Eqns.~\ref{eq5}. This gives the condition
\begin{equation}
\left(\lambda/n\right)^4+\left(\lambda/n\right)^2
+[3K-K^2-\Gamma^2]=0
\end{equation}

When the 3 bodies are slightly displaced from equilibrium,
they will undergo small oscillations about this point provided that
all of the frequencies are purely imaginary, and this requires
that
\begin{equation}
1\ge4(3K-K^2-\Gamma^2)\ge 0
\end{equation}

For the restricted case, where $m_2=0$, we recover the standard
condition for stability that
\begin{equation}
\frac{m_1}{\mtot}\le \frac{9-\surd 69}{18}=0.03852\ldots
\end{equation}
For the case of equal-mass secondaries, where $m_2=m_1$, we find
that the sum of their masses must be slightly less than the maximum
mass for the restricted case:
\begin{equation}
\frac{m_1+m_2}{\mtot}\le \frac{6-4\surd 2}{9}=0.03812\ldots
\end{equation}
It is straightforward to show that the critical value of
$(m_1+m_2)/\mtot$ always lies between these 2 values when
$0<m_2<m_1$. (After submitting the original version of this manuscript,
Michael Nauenberg informed us that Siegler and Moser (1971) 
derived an equivalent expression for the existence of periodic orbits in the
general 3-body problem).

For the range of masses in which the equilibrium points are stable,
$K$ and $\Gamma$ are small quantities, and we can derive approximate
expressions for the oscillations periods $P$, which are related
to the frequencies by $P=2\pi/|\lambda|$. The periods are given by
\begin{eqnarray}
P_1&\sim&P_{\rm Kep}\left[1+\frac{27(m_1+m_2)}{8\mtot}\right]
\nonumber \\
P_2&\sim&P_{\rm Kep}\,\sqrt{\frac{4\mtot}{27(m_1+m_2)}}
\end{eqnarray}
where $P_{\rm Kep}=2\pi/n$ is the orbital period. The first of
these is clearly close to the orbital period, while the second
represents the period of librations in $\theta$ about the
equilibrium point. For a pair of planets with masses comparable to
Jupiter or Saturn orbiting a solar-mass star, the libration period
is roughly an order of magnitude larger than the orbital period.

Figure 1 shows two examples of the behavior of equal-mass
planets with $m_{1}=m_{2}=0.25 M_{\rm JUP}$ which are orbiting a solar mass
central star and participating in co-planar co-orbital librations. 
In each case, the osculating initial
eccentricities and arguments of periapse are given by
$e_{1}=e_{2}=0.01$, and $\varpi_{1}=\varpi_{2}=0.0$. Planet 1
is started at periastron passage, and Planet 2 is started 60 days
ahead of periastron passage. 
The motion is plotted in the co-rotating frame over a 50 year timespan.
The first orbit has initial
osculating periods $P_{1}=365 {\rm d}$, and $P_{2}=355 {\rm d}$. In this case,
the planets exhibit Tadpole-like librations about the equilateral
configuration. The second orbit has a more discrepant pair of initial
osculating periods, with $P_{1}=370 {\rm d}$ and $P_{2}=350 {\rm d}$.
In this case, the larger initial perturbation from equilateral 
equilibrium causes the planets to execute horseshoe-like librations, and
the motion is symmetric about a $180^{o}$ separation. For both orbits,
there is no visible alteration in behavior over a 10 million year
trial integration.

We note that the Saturnian satellites Janus and Epimetheus provide
a solar system example of a horseshoe-type 1:1 resonance. Janus and
Epimetheus have a mass ratio $M_{J}/M_{E} \sim 0.25$. Their combined
mass is tiny in comparison to Saturn, however, $\mu=4.6\times10^{-9}$
(Yoder, Synnott, \& Salo 1989).

Figure 2 shows a series of surfaces of section in which the
difference in osculating semi-major axis ($a_{2}-a_{1}$) is plotted against
the longitudinal separation of the planets every time planet 1 reaches
periapse. For initial period differences $P_{1}-P_{2}<20 {\rm d}$, as
with the first example orbit shown in Figure 1,
the orbits are of the tadpole-type. A separatrix near $P_{1}-P_{2}=20 {\rm d}$
shows that there is a transition to horseshoe type orbits at larger
initial period separations. For the assumed initial conditions, the
last stable horseshoe orbit occurs when $P_{1}-P_{2}=27 {\rm d}$. Beyond this
initial period separation, the twice-per-libration
close encounters between the planets
impart momentum kicks that are large enough to disrupt the resonant
relationship. As the planet masses are increased, the width of the
stable horseshoe region in $\Delta P$ decreases. When the planets
reach $M_{1}=M_{2}=0.4 M_{\rm JUP}$, a set of numerical integrations shows that
the region of stable horseshoe orbits has disappeared. One can thus 
hope to find pairs of Saturn-mass planets executing horseshoe orbits around
a solar mass star ($M_{\rm SAT}/M_{\odot} \sim 0.0003$,
but one will not find two Jupiter-mass planets participating in this 
configuration ($M_{\rm JUP}/M_{\odot} \sim 0.001$).
Work is under way (Novak, Laughlin \& Chambers 2002) which
will further delineate the various stability boundaries for 1:1 configurations.

We point out that the surfaces of section corresponding to the
motion of equal-mass planets (Figure 2) produce slightly fuzzy curves which
are not clearly bounded. The finite width of these sections indicates
that there is no isolating integral constraining the motion. In the planar
restricted three-body problem, such an integral (the Jacobi constant, $C_{J}$),
does exist, and the analogous
curves in the surface of section have sharp boundaries. In this case, however, the
motion of the planets is not restricted, and over extremely long periods, the 
system should be able to diffuse out of the resonant configuration, unless
the orbits are strictly periodic.

Figure 3 shows the radial velocity component of
the central star along a line of sight which views the two systems
of Figure 1 in an edge-on
configuration. The time baseline is 15 years. Over this period, the planets 
undergo a significant fraction of a libration. The resonant motion is
visible in the stellar reflex velocity
as an envelope modulating the high frequency sinusoid corresponding to the
fundamental orbital frequency of the system. When the planets have
a large angular separation, their contributions to the radial velocity
of the star tend to cancel out. Furthermore,
the envelope modulation, with its
characteristic accordion-like shape is dictated by the masses of the planets,
and if a system is found in the 1:1 resonance, it
will not be subject to the $\sin i$ degeneracy
that pertains to Keplerian fits.

Figure 4 shows a fundamentally different 1:1 resonance that
may exist in extrasolar planetary systems. 
In this rather exotic configuration, one planet begins with a
high eccentricity orbit, while the other traces a lower
eccentricity orbit. We refer to this state of affairs as a
1:1 ``eccentric resonance''.
The configuration shown has equal-mass 1
$M_{\rm JUP}$ planets, on orbits with initial periods of $P_{1}=P_{2}=360 {\rm d}$. 
The planets initially avoid close encounters by maintaining relatively
small differences in longitude during pericenter passage of the
eccentric member of the pair. In the situation shown in Figure 4, at the moment of
periastron passage of the eccentric planet, the low eccentricity planet
has already passed periapse, and has a mean anomaly $M=40^{o}$. The
planet-planet perturbations are most effective when the eccentric planet
is near aphelion. During this period, which is marked by solid lines on 
the orbital curves, the planet on the near-circular orbit transfers
angular momentum to the eccentric planet.
As shown in Figure 5, this angular momentum transfer leads to a
periodic exchange of eccentricities between the planets over a $\sim 800$ year
secular cycle. (Heavier planets experience shorter secular cycles.)
Both planets experience only small amplitude high-frequency oscillations
in osculating semi-major axis during the secular cycle, indicating that
that the resonant lock is mediated by angular momentum rather than energy
exchange.
For $\mu=0.001$, the
resonant lock between the two planets is very stable, and is maintained
with no outward variation over the course of a ten million year test integration.
Numerical experiments show that the resonance breaks down for values 
$\mu \sim 0.035$.

As was the case with the tadpole and horseshoe type 1:1 resonances,
the line-of-sight radial velocity variation of the central star shows a distinctive
pattern, and is plotted over the initial ten year span in Figure 6.
Each periastron passage of the eccentric planet is marked by a steep
decrease in the radial velocity. Within a radial velocity data set, the presence
of an eccentric 1:1 resonance might typically be indicated by a single highly errant
velocity measurement superimposed on on otherwise sinusoidal pattern.

We note that because the 1:1 resonance is the lowest-order resonance, the effect
of planet-planet perturbations are seen most dramatically
in a radial velocity data set of given precision. Indeed, it may be the case
that
such systems are currently sitting unrecognized within the accumulated data
of the ongoing radial velocity surveys.

\section{Orbital fitting of 1:1 resonant systems}

As discussed above, and shown in Figures 3 and 6,
a pair of extrasolar planets participating in the 1:1 resonance produces
a readily detectable radial velocity signature. Because the
1:1 resonance involves strong planet-planet interactions, however, the usual
procedure of testing Keplerian planetary orbits to minimize the $\chi^{2}$
of the orbital fit will not give correct results.

The discovery of the 2:1 resonance between GJ 876 b and c has shown 
that planet-planet dynamical effects provide a crucial descriptive ingredient
of a particular system (Laughlin \& Chambers, 2001; Rivera \& Lissauer, 2001;
Nauenberg, 2002).
In this section we describe the application
of a straightforward self-consistent
dynamical fitting procedure which can uncover a 1:1 resonant configuration
within a set of radial velocities.

In order to illustrate the fitting procedure,
we have adopted the latest Keck Telescope radial
velocity data set for GJ 876 (Marcy et al 2001; Marcy, personal communication).
This data set includes 63 radial velocity measurements spanning 1532 days. 
Each velocity
has an associated measurement uncertainty which is estimated from the 
internal cross correlation
of the lines within the spectra. These velocity uncertainties range from 2.8
m/s to 8.3 m/s, and their r.m.s. value is 5.14 m/s. In Table 1, we have specified
the osculating orbital elements of a hypothetical tadpole-type resonant pair
orbiting a solar-mass star.
Using these orbital elements at the epoch of the first radial 
velocity point as initial conditions, we integrated
the hypothetical system forward in time for the duration spanned by the actual
radial velocity observations of GJ 876. We then sampled the radial velocity of
the parent star in response to the hypothetical 1:1 resonant pair at the 63
observational epochs. For each point, we then added radial velocity noise 
drawn from a
Gaussian distribution of half-width given by the actual quoted velocity errors.
This yields a synthetic radial velocity data set from which we can attempt to
reconstruct the orbital parameters of the planets.

The top panel of Figure 7 shows the
power spectrum of the synthetic radial velocity data set. There is a single
significant peak at the fundamental 30 day period, and the spectrum gives little indication 
of the presence of two planets in the data. In particular,
the accordian-like modulation of the radial velocity curve due to
the 300 day libration frequency between the planets
is not immediately apparent in the power spectrum, as can be seen by
comparing with the power spectrum produced by a single planet of
$K=75 m/s$ on a circular 30 day orbit (bottom panel of Figure 7).
In both cases, the power spectrum shows only a single significant peak.

In order to fit the synthetic 1:1 resonant
system, we have employed a two-stage method. In the
first stage, we use a genetic algorithm (Goldberg, 1989) as implemented in
Fortran by 
Carroll (1999) for public domain use. 
The genetic algorithm starts with an aggregate of osculating orbital elements,
each referenced to the epoch of the first radial velocity observation
($T_{0}=2450602.0931$). Each set of elements (genomes)
describes a unique three-body integration and an associated radial 
velocity curve for the central star. The fitness of a particular genome is
measured by the $\chi^{2}$ value of its fit to the radial velocity
data set. At each generation, the genetic algorithm evaluates the $\chi^{2}$
fit resulting from each parameter set, and cross breeds the best members of the 
population to produce a new generation.

Because the fundamental 30 day period of the system is clearly visible
in the power spectrum,
the genetic algorithm is constrained to search for 1:1 resonant configurations
in which the initial period ranges for the planets are $29 {\rm d} < P_{1} < 31 
{\rm d}$, and $29 {\rm d} < P_{2} < 31 {\rm d}$. The initial arguments of
periapse and mean anomalies of the two planets are allowed to vary within
the allowed $2\pi$ range. The radial velocity half-amplitudes of the planets
are required to fall in the range $0<K_{1}<150 {\rm m/sec}$ and
$0<K_{2}<150 {\rm m/sec}$. The planetary eccentricities are allowed to vary within
the full $0<e<1$ range.

With these very liberal constraints on the space of osculating initial
orbital elements, the Genetic Algorithm rapidly identifies
a set of parameters having $\chi=1.27$. This tentative fit is then
further improved by use of Levenberg Marquardt minimization (Press et al.
1992) to produce a fit to the data having $\chi=0.97$. This fit is
shown in Figure 8, and in the third and fourth columns of Table 1.
The excellent agreement between the input system and the fitted solution 
shows that a 1:1 resonant pair is readily identifiable if it exists
within a radial velocity data set comparable to the GJ 876 velocities. 
As the radial velocity surveys continue, an increasing number of target
stars will have velocity data sets of this quality.

\section{The Formation and Migration of Trojan Planets}

The foregoing sections show that 1:1 resonant configurations are viable from
an orbital stability standpoint, and are also readily detectable should
they exist. The natural question, therefore, is: Can co-orbital planets
form? We believe that the answer is yes, and in a forthcoming work, we
will present a detailed argument in favor of this hypothesis (Novak,
Laughlin, \& Chambers 2002). Here, we briefly outline our arguments.

The parameter space available to the eccentric 1:1 resonance and its
close relatives has not yet been mapped in detail, but initial numerical
surveys show that it is surprisingly large. A vast range of co-orbital
configurations exist which are both stable and which roughly resemble
the intermediate panels of Figure 5. It is challenging to imagine how
such systems might form, but one scenario is as follows:
A planetary system forms
in which there are three major planets. The $\upsilon$ Andromedae system
provides a rough example (Butler et al 1999). A dynamical instability
or collision then occurs between two of the planets, leaving two survivors
participating in some form of the eccentric 1:1 resonance. We would expect 
that such dynamical formation channels are unlikely, but
the cross section for this class of events could be calculated numerically
in the event that a system of this type is discovered
(Laughlin \& Adams 1998, Ford, Rasio, \& Havlikova 2001).

We expect that 
Tadpole or Horseshoe-type planetary pairs are more common, and that
their formation is
mediated by interaction with a protostellar disk. One possibility
is that a pair of planets can form directly in a near equilateral
configuration. Figure 9 shows a detail of a hydrodynamical calculation
which was performed and discussed by Laughlin, Chambers and Fisher (2002).
In this scenario, a planet with mass $0.75 M_{\rm JUP}$ is present within
a protoplanetary accretion disk. The viscosity parameter of the disk
in the region near the planet is of order $\alpha \sim 0.02$, and the
planet, with its sub-Jovian mass, is only marginally capable of maintaining
a gap in the disk. Material is seen to linger in the vicinity of the 
$L_{4}$ and $L_{5}$ points. The vortical flow pattern in these regions
might lead to particle trapping and the accumulation of a second significant
core (e.g. Adams \& Watkins 1995, Barge \& Sommeria 1995, Godon \&
Livio 2000). Recent high-resolution numerical simulations by 
Balmforth \& Korycansky (2001) have shown that the large-scale vortical
flow enveloping the horseshoe region nucleates additional smaller 
vortices, which may spur further formation of planetary cores.

If two planets form from a disk in a 1:1 horseshoe or tadpole configuration,
the resonant lock is maintained as the planets migrate inwards as a result
of their interaction with the disk. This effect is shown in Figure 10,
which shows the result of a three-body integration of a star and a resonant
1:1 tadpole type Jovian-mass planetary pair. A small azimuthal torque
$a_{\phi}=-3.4 \times 10^{-5} {\rm AU/ yr^{2}}$ is applied to one of
the planets. Despite this uneven application of torque, the planets remain
coupled together in resonance, and migrate inwards together. 
Resonant migration leads to secular eccentricity increase
for planets on first or higher order mean motion resonances, and as
noted by Lee \& Peale (2002), this increase leads to difficulties in
constructing a plausible formation scenario for the planets orbiting GJ 876.
In the case of 1:1 resonant migration, however, angular momentum and energy
loss from the planets occurs at a rate which prevents a secular increase
in eccentricity. Once formed, a 1:1 resonant pair is capable of migrating
to small semi-major axes.

In our upcoming paper, we will describe a sequence of events which we
believe is the most likely channel for forming observable 1:1 configurations.
Our scenario stars with a major collision between two multiple Earth-mass
cores in a nascent planetary system. The collision leads to the formation
of a binary core, in much the same fashion as a giant impact is believed
to have formed the Earth-moon system. The cores retain their individuality,
and accrete gas from the surrounding nebula, leading to the formation of
a bound double planet. This double planet migrates inward to small
orbital radii, at which point the separation of the planets exceeds the
Hill Radius. At this point, the planets scatter onto a horseshoe-type
orbit, which can be subsequently damped to a trojan configuration.

\subsection{Acknowledgments} 

We would like to thank Debra Fischer, Doug Lin, Geoff Marcy, Michael
Nauenberg, and Greg Novak for useful discussions. Stan Peale provided a very
prompt and informative referee's report.
This material is based upon work supported by NASA under contract No.
RTOP 344-37-22-12 issued through the Origins of Solar Systems Program, 
and by funds from the University of California, Santa Cruz. JEC
acknowledges support from an NRC Postdoctoral Program and from the
NASA Origins of Solar Systems Program.

\clearpage

\begin{figure}
\caption{
Examples of trojan orbits and horseshoe orbits. A detailed plot is
shown in the right panel.
In each case, the mass of both planets is 1 $M_{\rm JUP}$.
The solid line corresponds to the horseshoe-type orbit. The
dotted line corresponds to the tadpole-type orbit.
}
\end{figure}

\begin{figure}
\caption{
Surface of section showing $a_{1}/a_{2}$ and the difference in longitude
$D=(M_{1}+\varpi_{1})-(M_{2}+\varpi_{2})=\lambda_{1}-\lambda_{2}$ 
for a pair of co-orbital
planets every time planet 1 reaches periapse. The loci of points centered
on $a_{1}/a_{2}=1.0$, $\lambda_{1}-\lambda_{2}$=1.0472 are executing 
tadpole-type orbits. The outermost loci correspond to horseshoe-type orbits.
}
\end{figure}

\begin{figure}
\caption{
Synthetic radial velocity variations for the planetary systems shown
in Figure 1. The solid line corresponds to the horseshoe-type orbit.
The dotted line corresponds to the tadpole-type orbit. 
}
\end{figure}

\begin{figure}
\caption{
Planet orbits in an example 1:1 eccentric resonance. The solid lines
indicate portions of the respective orbits swept out in a single period
of time corresponding to 15\% of the orbital period.
}
\end{figure}

\begin{figure}
\caption{
Evolution of planet orbits in an example 1:1 eccentric resonance.
}
\end{figure}

\begin{figure}
\caption{
Synthetic radial velocity variations for a planetary system participating
in the 1:1 eccentric resonance. The radial velocity curve corresponds
to the situation near the first panel of Figure 5, where one planet
is highly eccentric while the other is on a near-circular orbit.
}
\end{figure}

\begin{figure}
\caption{
Upper panel: Power spectrum corresponding to a synthetic radial
velocity data set produced using the hypothetical planetary system
shown in Table 1, sampled at the epochs and velocity precision of
the GJ 876 Keck radial velocities. Lower Panel: Power spectrum
corresponding to a single planet on a circular orbit with $K=75 {\rm m/s}$,
and $P=30 {\rm d}$, also sampled at the epochs and velocity
precision of the GJ 876 data set.
}
\end{figure}

\begin{figure}
\caption{
Synthetic radial velocity variations for the planetary system shown
in Table 1 (solid line). The circular points with small vertical lines corresponding to
 errorbars represent a sample
of the system with the properties of the GJ876 data set.
A superimposed dotted line
line shows a fit to the sampled data obtained with a combined
Genetic Algorithm and Levenberg-Marquardt procedure. It is hard to
distiguish the difference between these two curves at the resolution
of the plot.
}
\end{figure}

\begin{figure}
\caption{
Detail of a hydrodynamical calculation of a planet clearing a gap
in a disk. A vortical flow is observed around the L5 point. We hypothesize
that such a vortex could trap particles and lead to the formation of
a second planet trapped in 1:1 resonance with the first.
}
\end{figure}

\begin{figure}
\caption{
Evolution of a three-body simulation in which a pair of equal
mass planets in 1:1 resonance migrate inward without showing secular
increase in eccentricity. Torque is applied to only one of the planets.
In the bottom panel, the eccentricities of both planets are shown,
with planet b's eccentricity plotted as dots, and planet c's eccentricity
plotted as a solid line.
}
\end{figure}

\newpage

\begin{table}
\begin{center}
\begin{tabular}{lcccc}
\hline
\hline
\\
 Parameter &  Planet 1 & Planet 2 & Planet 1 & Planet 2 \\
           & \multicolumn{2}{c}{(Assumed Elements)} & 
             \multicolumn{2}{c}{(Fitted Elements)} \\
\\
\hline
\\
 Period (day)          & 30.000  & 30.000 & 29.919 & 30.037 \\
 $K$ (ms$^{-1}$)       & 50.000  & 100.00 & 49.233 & 99.694 \\
 Eccentricity          & 0.0500  & 0.0500 & 0.0418 & 0.0401 \\
 $\omega$ (deg)        & 180.00  & 180.00 & 195.02 & 174.29 \\
 Periastron Time (JD)  & 10.000  & 20.000 & 11.253 & 19.512 \\
\\
\hline
\end{tabular}
\end{center}
\caption{\bf Assumed and Fitted Elements for 1:1 Resonant System
( using GJ 876 data set)}
\end{table}

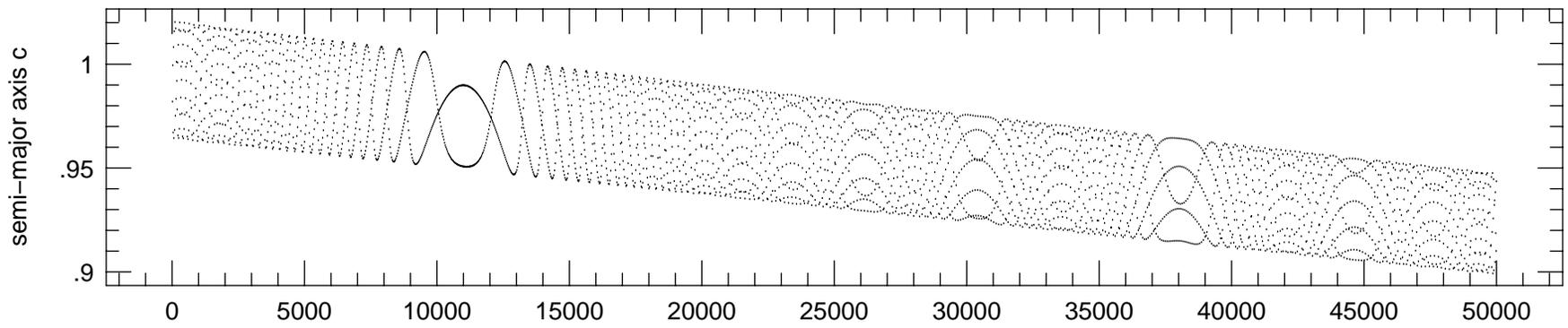
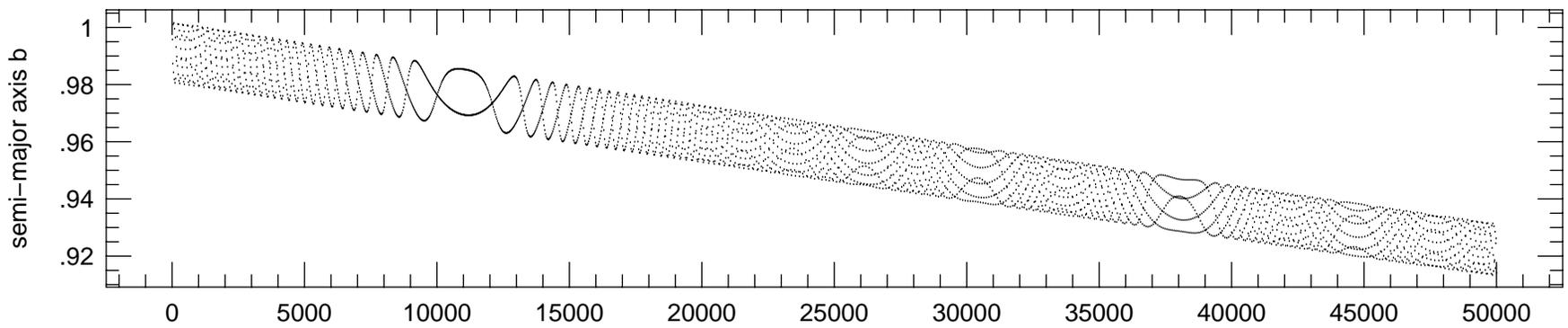
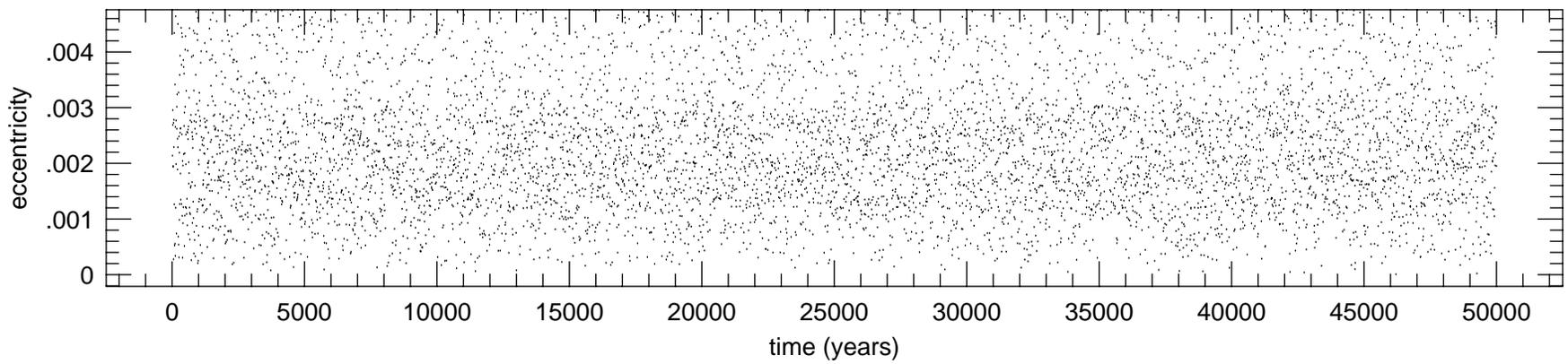
\end{document}